\documentclass[12pt]{article}
\usepackage[dvips]{color}
\usepackage[dvips]{graphics}
\usepackage[T1]{fontenc}
\usepackage[latin1]{inputenc}
\usepackage{longtable}
\usepackage{times}
\usepackage{bm}

\usepackage[normalem]{ulem}
\usepackage{geometry}
\newcommand\suppress[1]{}

\newlength\wvtextpercent
\newbox\strikebox
\def\strike#1{\setbox\strikebox \hbox{<#1>}\hbox{\raise0.5ex\hbox to 0pt{\vrule height 0.4pt width \wd\strikebox\hss}\copy\strikebox}}

\def\la{\mathrel{\mathchoice {\vcenter{\offinterlineskip\halign{\hfil
$\displaystyle##$\hfil\cr<\cr\sim\cr}}}
{\vcenter{\offinterlineskip\halign{\hfil$\textstyle##$\hfil\cr
<\cr\sim\cr}}}
{\vcenter{\offinterlineskip\halign{\hfil$\scriptstyle##$\hfil\cr
<\cr\sim\cr}}}
{\vcenter{\offinterlineskip\halign{\hfil$\scriptscriptstyle##$\hfil\cr
<\cr\sim\cr}}}}}
\def\ga{\mathrel{\mathchoice {\vcenter{\offinterlineskip\halign{\hfil
$\displaystyle##$\hfil\cr>\cr\sim\cr}}}
{\vcenter{\offinterlineskip\halign{\hfil$\textstyle##$\hfil\cr
>\cr\sim\cr}}}
{\vcenter{\offinterlineskip\halign{\hfil$\scriptstyle##$\hfil\cr
>\cr\sim\cr}}}
{\vcenter{\offinterlineskip\halign{\hfil$\scriptscriptstyle##$\hfil\cr
>\cr\sim\cr}}}}}

\begin{document}
\sloppy

\textbf{The Physics of E x B-Drifting Jets}

\textbf{}

Wolfgang Kundt$^1$ and Gopal Krishna$^2$

\noindent\textit{$^{1)}$ }\textit{IfA of Bonn University, Auf dem Hügel 71, D-53121 Bonn, wkundt@astro.uni-bonn.de}

\noindent\textit{$^{2)}$ }\textit{National Centre for Radio Astrophysics/TIFR, Pune University Campus, Pune-411007, krishna@ ncra.tifr.res.in}

\textbf{}

J. Astrophys. Astr. (2004) \textbf{25}, 115-127

\textbf{}

\textbf{Abstract.   E} x \textbf{B}-drifting jets have been generally ignored for the past 25 years even though they may well describe all the astrophysical jet sources, both on galactic and stellar scales. Here we present closed-form solutions for their joint field-and-particle distribution, argue that the observed jets are near equipartition, with extremely relativistic, mono-energetic $e^{\pm}$-pairs of bulk Lorentz factor $\gamma \la 10^4$, and are first-order stable. We describe plausible mechanisms for the jets' (i) formation, (ii) propagation, and (iii) termination. Wherever a beam meets with resistance, its frozen-in Poynting flux transforms the delta-shaped energy distribution of the pairs into an almost white power law, $E^2N_E \sim E^{-\epsilon}$ with $\epsilon \ga 0$, via single-step falls through the huge convected potential. 

\textbf{}


\textbf{}

\textbf{\hfill{}1. Background\hfill{}}

\textbf{}

Pair-plasma jets with ultra-relativistic bulk motion have been proposed twenty-five years ago by one of us (Kundt, 1979), then jointly elaborated by us (Kundt \& Gopal-Krishna, 1980). They were also proposed by Morrison (1981), but have usually not been mentioned (cf. Begelman et al, 1984, 1994). Despite intermediate progress reported, e.g., in Kundt \& Gopal-Krishna (1986), Blome \& Kundt (1988), Baumann (1993), and in Kundt (1996, 2004), they have been treated with a healthy scepticism by the community, cf. Blandford (2001), and Beresnyak et al (2003), with a few notable exceptions, e.g. Reipurth \& Heathcote (1993), Scheuer (1996), Prieto et al (2002), Brunetti (2002), and Stawarz (2003). A possible reason for this lack of widespread acceptance may have been a concern about whether or not the beams allowed a stable transport of a broad energy distribution of high-energy charges, in the form of an ordered \textbf{E} x \textbf{B}-drift. The beams are indeed unlikely to transport a broad distribution. 

   Instead, their Poynting-flux-flooded formation regions are expected to generate particle distributions at least as sharp in 4-momentum as relativistic Maxwellians, and an onsetting      \textbf{E} x \textbf{B}-drift will further sharpen the narrow distribution towards a delta-type one. Such equipartition pair-plasma flows convect half of their energy as a stationary Poynting flux which is ready -- wherever stalled -- to broaden the particle distribution into an almost white power law, starting with Lorentz factors of order $10^2$ at their bottom end, and extending up to Lorentz factors of order $10^6$, in the form of a long high-energy tail whose radiated power peaks at the top end whereas its energy density peaks near its bottom end. During undisturbed propagation, such beams are loss-free on Mpc scales except for minor inverse-Compton losses on the radiation background. Note that peripheral tapping of a beam can reveal its monoenergetic distribution, as in Sgr A*: the convected fields vanish at the channel wall. 

   We shall present exact solutions for such monoenergetic beams in section 2, and show that they are stable to first order. In the two subsequent sections, we shall offer reasonings why such beams are expected to form naturally around magnetized rotators, i.e. around rapidly rotating stars as well as in the centers of galactic disks, and why their radiation is expected to take the form of broad power-laws, from the radio to the X-ray and gamma-ray regime, with certain  emission dips and excesses which can be understood as due to strong anisotropies in their emission patterns at high frequencies. This uniform model covers the observed jets from (a) newly forming stars (or YSOs), (b) forming white dwarfs, inside planetary nebulae (PNe), (c) young binary neutron stars (within light or heavy accretion disks), and from (d) the nuclear-burning centers of galactic disks (or AGN). 

Note that monoenergetic relativistic electron beams have just been produced in the lab: Katsouleas (2004).  A few extreme and/or controversial jet sources are discussed in section 5.

\textbf{}

\textbf{\hfill{}2.  Solving the Beam Equations\hfill{}}

\textbf{}

As has already been argued -- and will be elaborated in the next two sections -- the beams of the jet sources are expected to consist of overall electrically neutral and current-free configurations of electrons and positrons at large Lorentz factors $\gamma > 10^2$, convecting toroidal magnetic fields and `radial' electric Hall fields w.r.t. the roughly cylindrical geometry of a beam segment. In reality, such segments of stationary flow have an approximately conical shape, but will be approximated by us, for simplicity of presentation, by cylinder segments. Note that even a conical beam does not have (adiabatic expansion) losses when moving through a (strictly) vacuum channel.

We describe a beam segment by cylindrical coordinates {z, s, $\varphi$}, with z growing parallel to the beam axis, and s, $\varphi$ being polar coordinates in the cross-sectional planes. In these coordinates, Maxwell's (stationary, axially symmetric) equations  $\bm{\nabla} \cdot \textbf{E }= 4\pi \rho$ , $\bm{\nabla} \times \textbf{B} = (4 \pi /c) \textbf{j }$, $\bm{\nabla} \times \textbf{E} = 0 = \bm{\nabla} \cdot \textbf{B}$  yield respectively:

\textbf{ }

\hfill{}$\partial _s (s E_s )  =   4 \pi s \rho$\hfill{}(1)

\textbf{ }

\hfill{}$\partial_s (s B_{\varphi})  =  4 \pi s \rho \beta_z$   ,   $\partial_s B_z  =  - 4 \pi \rho \beta_{\varphi}$ \hfill{}(2)

\textbf{ }

\noindent with  $\bm{\beta}$ := \textbf{v}/c , and with all quantities only depending on the radial coordinate s. We restrict our attention to  \textbf{E} x \textbf{B}-drifting charges, for which the acceleration

\textbf{ }

\hfill{}$c(\gamma \bm{\beta})^{\bm{\cdot}}  =  (e/m_e) (\textbf{E} + \bm{\beta} \times \mathbf{B})$\hfill{}(3) 

\noindent vanishes, yielding 

\hfill{}$E_s  =  B_{\varphi} \beta_z  - B_z \beta_{\varphi}$ .\hfill{}(4)

\textbf{ }

   For realistic solutions with net charge and current zero, we have the additional boundary conditions at  s = 0 and R (:=  beam radius):

\hfill{}$(s E_s)(0)  =  0  =  (s E_s)(R)$ ,\hfill{}

\hfill{}\hfill{}(5)

\hfill{}$(s B_{\varphi})(0)  =  0  =  (s B_{\varphi})(R)$ ,\hfill{}

\textbf{ }

\noindent and the general stationary, cylindrically symmetric solution can be Fourier-expanded w.r.t. s/R:

\textbf{ }

\hfill{}${s E_s , s B_{\varphi}}$  $\sim$  $\Sigma_{k \ge 1} C_k sin(k \pi s/R)$ ,\hfill{}(6)

\textbf{ }

\noindent of which the first (ground) term already contains most of the physical information. 

   For large $\gamma$ (>$10^2$), we have $\vert \beta_{\varphi} \vert << \beta_z  \approx 1$, and the leading term of the expansion reads:

\textbf{ }

\hfill{}$E_s \approx B_{\varphi} \approx  C sin(\pi s/R) / s$  ,  $\rho \approx (\pi C/R) cos(\pi s/R) /s \approx  j_z /c$  ,  $B_z \approx const$ .\hfill{}(7)

\textbf{ }

\noindent It expresses a uniform flow in z-direction, with a positive \{or negative\} net charge density inside of R/2, and the opposite charge density dominating for s > R/2 , and correspondingly with a net \{positive, negative\} current density \{inside, outside\} of  s = R/2 , both of which peak at both ends, on the axis as well as at the periphery; see Fig. 1. 

\begin{figure}
\centering
\resizebox{070mm}{!}{\includegraphics{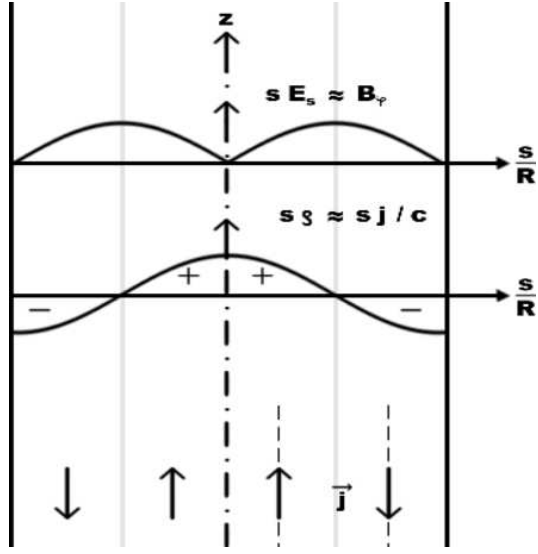}}
\caption{ Cross section through (the ground mode of) a beam segment, showing the radial
dependences of $\rho, j, E_s , and B_{\varphi}$.
}
\label{Figure1}
\end{figure}

   This solution is as simple and transparent as one could have imagined; the involved charge densities $\rho$ are a tiny fraction of those composing the beam: $\rho /e n_e = 10^{-9.2} \gamma _3 /L_{44}^{1/2}$, i.e. correspond to a minute distortion of charge neutrality. For a jet of power L and cross-sectional area A , (ram) pressure equipartition (among particles and fields) requires

\textbf{ }

\hfill{}$\gamma n_e m_e c^2   =  (E^2 + B^2)/8 \pi \approx B^2/4\pi \approx L/A c$\hfill{}(8)

\textbf{ }

\noindent where $n_e$ is the electron number density. (Equipartition is plausible from what we have said, and conforms with the observations, cf. Begelman et al, 1984). Note that the convected electric potential  $\Phi = \int E_s ds$  for typical jets can be gigantic, of order  $e \Phi \approx e(\pi L/c)^{1/2}  =  10^{19.5}$eV $L_{44}^{1/2}$ ; we shall see below that it can generate a high-energy tail reaching up to electron Lorentz factors of $10^6$ and more. Note also that $\Phi$ is scale-invariant: $A$ has dropped out, via $s^2$; $\Phi$ depends solely on the source power $L$, which is still large for stellar jet sources, with $L \approx 10^{35 \pm 1}$erg/s.

   Realistic jets have $\beta_z < 1$ and $\beta_{\varphi}\neq 0$. For them, a straight-forward calculation, starting from equ. (4) and using the integrated equs. (1,2), leads to

 \textbf{ }

\hfill{}$\rho \beta_{\varphi} = \partial _s[\int ds s ^{- 2}{\partial
_s(\int ds \rho s)^2  -  \partial _s (\int ds \rho  \beta _z
s)^2}]^{1/2}  \approx  \partial _s [\int ds s ^{- 2}  \partial _s (\int
ds \rho s)^2]^{1/2}/ \gamma$ ,\hfill{}(9)

\textbf{ }

\noindent the latter for  $\beta_z \approx 1 - 1/2 \gamma^2$. It shows that  $\beta_\varphi$ is small of order  $1/\gamma$ , i.e. that ordered spiralling of the charges should be unimportant in high-energy jets. This result must not be confused with the existence of helical beams, in interaction with turbulent (heavy) environs.

   So far, we have assumed strictly monoenergetic beams, with $\beta_z =
const$, which cannot be expected under realistic conditions. Charges
whose $\beta_z$ deviates from (the local value of) $E_s/B_{\varphi}$ will
violate equ. (4), and start moving radially inward or outward,
depending on their sign, whereby both \{hard, soft\} charges move
\{inward, outward\} for opposite signs. In each case, a glance at equ.(3)
shows that the radially moving charges of deviant $\gamma$ must
fall through the electric potential $\Phi$ such that their 4-momentum
is adjusted to the (locally) appropriate value for a stationary
drift: more energetic ones lose, less energetic ones gain in energy,
independent of the sign of their charge. This key stability is
intuitive already on energetic grounds, from the shape of $\Phi$, but
follows directly from (3) because any radial drift implies an
acceleration in $\pm z$-direction, via the $e(\textbf{E} + \bm{\beta}
\times \textbf{B})$-term (whose sign changes with the sign of $e$, and
likewise with the sign of $\bm{\beta}$ $- <\bm{\beta}>$).

   Note that in principle, $\beta_z$ could have been radius-dependent so that a finite spread in $\gamma$-s is transported by the beam. But such a fine-tuning of fields and particles is unlikely to be stable, after what we have just found: The wings of a distribution (in $\gamma$) are removed during short distances via radial falls (of proper sign) through the convected \textbf{E}-fields -- short of order $10^{-10}$ R because the \textbf{E}-fields are gigantic, as shown below equ. (8) -- so that individual particle energies are stabilized quickly.    

\textbf{}

\textbf{\hfill{}3.  Forming the Jets\hfill{}}

\textbf{}

We shall now argue that monoenergetic beams -- as have just been considered -- are expected for the cosmic jet sources, i.e. are not grossly overidealized.

   To begin with, there has to be an abundant source of relativistic $e^{\pm}$ : Magnetic reconnections, leading to cavities with dominating Poynting flux (because weak fields would simply be anchored by the charges, hence would not decay) are familiar from the solar surface, and are correspondingly expected near the inner edges of stellar accretion disks, and near the innermost, strongly shearing galactic disks, in scaled-up proportions (Kundt, 1996, 2002). Such coronal magnetospheric reconnections in non-rigid rotators are expected to involve comparable powers to thermal emissions -- cf. the magnetoid model of Ozernoy \& Usov (1977) -- because they drain on comparable energy reservoirs (controlled by equipartition). They can easily fill up the local hot bubble seen as the Broad Line Region (BLR) in AGN sources, which discharges to both sides of the disk through Blandford and Rees's (1974) deLaval nozzles, in the form of a supersonic twin-jet (Kundt, 1996), see Fig. 2.  

\begin{figure}[h]
\centering
\resizebox{100mm}{!}{\includegraphics{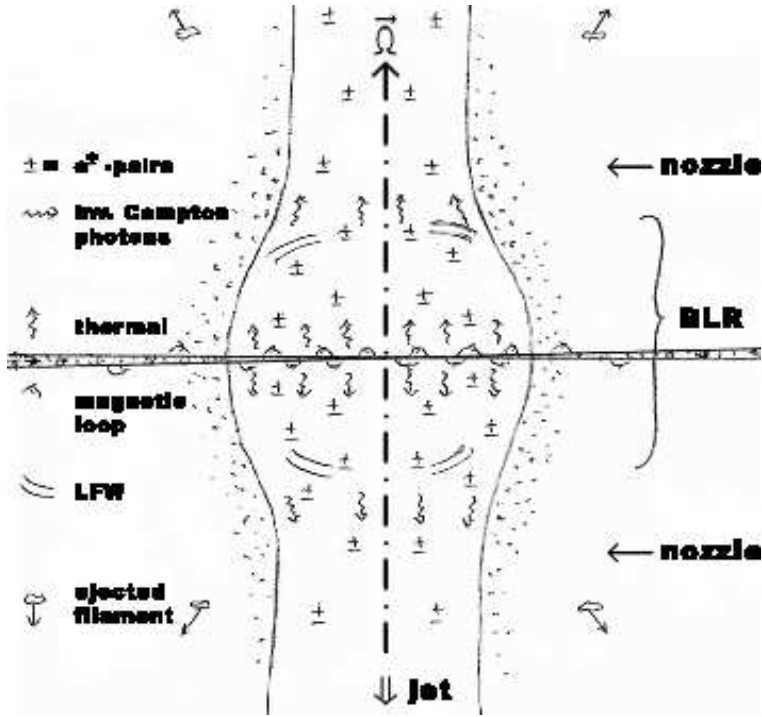}}
\caption{ 
Sketch of a plausible Central Engine: Coronal Magnetic Reconnections create $e^{\pm}$-pairs, the warm Central (Star and/or) Disk emits photons, Low-Frequency Waves post-accelerate the escaping $e^{\pm}$, and the latter boost the thermal photons to high-energy $\gamma$-rays. An ambient thermal bulge serves as the deLaval nozzle from which a twin jet emerges, along the spin axis of the central rotator; in galactic-center sources, this region is observed as the BLR.
}
\label{Figure2}
\end{figure}

   Once we deal with a central hot bubble filled with relativistic pair plasma, what will its energy distribution be like? In the laboratory, atomic beams are routinely cooled by shooting at them with a laser beam. In the BLR, the charges are post-accelerated by the simultaneously generated Low-Frequency (LF) waves of the central magnetized rotator, of angular frequency $\omega$ reaching down to some $10^{-4}s^{-1}$, hence of (large) strength parameter f :

 \hfill{}$f := eB / m_e$ c $\omega = 10^{14.2} B_3/\omega_{-4}$ \hfill{}(10)

\textbf{ }

\noindent for a typical coronal field strength B measured in KG; with $B_3 := B/10^3 G$, $\omega _{-4} := \omega /10^{-4} s^{-1}$. From the windzones of pulsars (like the Crab) we believe to have learned that the LF waves sweep the charges up in energy to Lorentz factors $\gamma$ of order $f^{2/3} = 10^{9.5}(B_3/\omega _{-4})^{2/3}$, in the absence of damping (Kulsrud et al, 1972; Kundt, 1986). 

   Damping occurs in the BLR through the equally present thermal (HF) radiation, a narrow `bump' between IR and X-rays, via inverse-Compton losses which truncate a distribution towards high energies, because they scale as $\gamma^2$. For class (d) (of AGN), these inverse-Compton losses are so strong that only some 10\% of them are radio-loud, and show jets. 

   As the outgoing charges in the (Thomson-opaque) BLR interact with both the LF and HF photons, their momentum distribution (away from the central engine) is expected to evolve towards a relativistic Maxwellian in radial direction, or even sharper, of Lorentz factor $\gamma  > 10^2$, (corresponding to brightness temperatures in excess of $10^{12}$K). Such high Lorentz factors are (i) expected, after equ. (10), are (ii) indicated by the overall energetics, (iii) by an avoidance of the inverse-Compton catastrophe, (iv) by the statistics of superluminal speeds (Kundt, 2004), and (v) by LF intraday variability, cf. Singal \& Gopal-Krishna (1985), Wagner \& Witzel (1995), but also Jauncey et al (2003) who prefer an interpretation via interstellar scintillations. 

   When the charges leave the BLR and approach the deLaval nozzle -- formed naturally by the obstructing plasma of the ambient circumstellar medium, or central galactic bulge respectively, whose inertia (in pressure balance) scales inversely as its temperature T, i.e. is $10^{8.3}/T_4$-times larger -- their narrow momentum distribution will be channeled into a monoenergetic one, as inferred above from equ.(3), with dominating Poynting flux which carries at least half the energy.

\textbf{ }

\textbf{\hfill{}4.  Discharging the Jets\hfill{}}

\textbf{ }

Once the highly-relativistic $e^{\pm}$-pairs from the BLR enter a vacuum channel, rammed by preceding generations of charges, they form an almost loss-free, monoenergetic \textbf{E} x \textbf{B}-drifting beam  as calculated explicitly in section 2  whose only losses are inverse-Compton collisions on the radiation background obeying:

\hfill{}$l_{deg} :=  \gamma /\gamma '  =  3 m_e c^2 / 4 \sigma _T u_{3K} \gamma$  =  Mpc / $\gamma _6 (1+z)^4$  \hfill{}(11)

\textbf{ }

\noindent where $l_{deg}$ is the degradation e-folding length, $\sigma _T$ the Thomson cross section, $u_{3K}$ the energy density of the 3K background radiation, and $\gamma _6$ := $\gamma /10^6$ , (Kundt, 2001). For a bulk Lorentz factor $\gamma \la 10^4$, thought to be realistic for most jet sources, inverse-Compton losses on Mpc scales are therefore ignorably small at redshifts z < 2.    

   Note that a monoenergetic beam has no collisional losses between its member charges -- because they have vanishing relative velocities -- nor dynamic-friction losses, for the same reason. Such internal-friction losses would in any case be ignorably small, because of the beam's extremely small electron-number density:

\hfill{}$n_e  =  L / A \gamma m_e c^3  =  10^{-8} $cm$^{-3} (L / A \gamma)_{-3.5}$ .\hfill{}(12)

\textbf{ }

\noindent During its propagation through a vacuum channel, a monoenergetic beam has only the inverse-Compton losses described by equ. (11).   

   Conditions change when a beam encounters obstacles, in the form of (heavy) channel-wall material, or channel intruders, or obstructing material at its downstream end, its `head'. Such obstructing plasma tends to be highly conductive, hence forbids penetration of electric and magnetic fields. The guiding toroidal magnetic field then gets compressed like the windings of a coil, and so are the convected charge clouds. Both electric charges and currents pile up against such a conducting wall, changing the field geometry in a way conveniently to be described by mirror charges (of opposite sign), and mirror magnetic fields (of same sign), see Fig. 3.

\begin{figure}[h]
\centering
\resizebox{100mm}{!}{\includegraphics{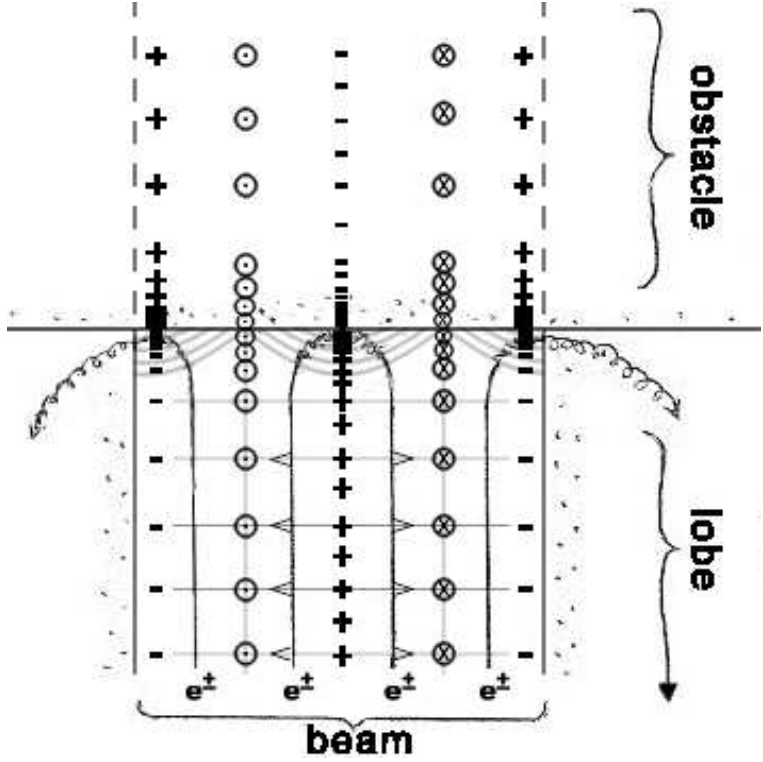}}
\caption{ 
Simplified cross section through a beam head, sketching the distributions of relativistic electrons ($e^{\pm}$), electric and magnetic fields, mirror charges and fields (on the side of the obstructing ambient plasma), and particle orbits. Omitted are the motions of the stalled charges escaping from the impact center, whereby they are post-accelerated by the huge convected potential.  
}
\label{Figure3}
\end{figure}

   Note that a charge-symmetric beam cannot be arrested, or reflected by electric fields alone; the latter can only redistribute the energies among the charges. The reflection of a (neutral) beam at its head is achieved by a changing geometry of both fields, electric and magnetic, such that the formerly quasi-stationary \textbf{E} x\textbf{ B}-drift in forward direction is diverted, partially towards the beam's axis, and partially sideways towards its periphery (Fig.3). At the same time, the bulk speed of the charges is reduced, from supersonic to subsonic, whereby straight-line motions change into gyrations. But now the charges are post-accelerated by the huge, convected electric potential, in the form of a space-charge limited flow whose relativistic version was first treated by Michel (1974), in application to pulsar polar-cap discharges.

   Michel's derivation of the relativistic generalization of Child's Law restricts itself to stationary, one-dimensional, one-fluid discharges inside a low-density plasma whose asymptotic speed is extremely relativistic, and whose asymptotic charge density realizes force-freeness, also known as Goldreich-Julian (1969) density, or Hones-Bergeson (1965) density. The derivation takes care of the fact that due to distributed screening, individual charges fall only through a tiny fraction of the available potential $\Phi$. This fraction, of order the square root of $e \Phi$ (in units of the electron rest energy), results as a consequence of Maxwell's equations plus conservation of energy, by integration along particle orbits through the (magnetized) polar gap:

\hfill{}$\gamma _{\infty} \approx  (8 \gamma_{\Phi})^{1/2}$  .\hfill{}(13)

\textbf{ }

\noindent The electron Lorentz factor $\gamma_{\infty}$ is reached exponentially towards the boundary of the polar gap, whose voltage $\Phi$ is assumed stabilized by unipolar induction.

   In the present case of a stalled beam, the convected potential $\Phi$ has its peak near the impact center, somewhat enhanced by compression w.r.t. its convected value, together with a surrounding ring wall of opposite sign. The boosting arena is therefore neither 1-d, 1-fluid, nor stationary. Still, distributed screening will result in the square of the charges' Lorentz factor $\gamma$ tending (as above) towards some multiple of the maximum available one, $\gamma_{\Phi}$ , and we expect the energy distribution in the stalled beam to acquire a high-energy power-law tail -- after averaging over a spatial ensemble of discharges -- reaching up from its convected value ($\gamma _{beam} \la 10^4$) all the way to its peak value, of order ($\gamma _{\Phi})^{1/2} \la  10^7$ , see equ. (8). At the same time, the charges of inappropriate sign (and same instantaneous flow direction) are decelerated to lower energies, extending the spectrum downward in energy below $\gamma_{beam}$. This sudden change of the energy distribution, from delta-like to hard power-law, takes place wherever a beam is stalled by ambient plasma, thanks to its convected Poynting flux.

   In this way, an almost loss-free, monoenergetic beam gets radiative whenever obstructed, with a broad power-law spectrum. A broad power law need not form, however, for peripheral tapping, in boundary-layer interactions.

   A well-known phenomenological dichotomy among the jet sources relates to the location of their hotspots, and tends to be called by their Fanaroff and Riley class I or II. Eilek et al (2002) have recently revised this classification into type A and type B, whereby class II is a subcase of type A, and speak of "straight" and "tailed" sources in the two cases. All jet sources start out "straight", their heads ramming supersonically into their ambient medium. During growth, the head's ram pressure drops as $r^{-2}$ with distance $r$ from the central engine, and the sound speed of the ambient medium often rises, so that beyond a certain distance -- which differs from source to source -- the head speed must pass from supersonic to subsonic w.r.t. the medium. From then on, the charges entering the terminating hotspot are no longer reflected (into their lobe) but continue coasting subsonically, in a gyrating mode, and can form a long, radiative "tail" (Gopal-Krishna et al, 1988, 1996). 

   In our understanding, this sonic transition marks the transition from Eilek type A to B. Beyond the decelerating hotspot, particle motions are no longer channelled, or lossfree. The stalled jet material "engulfs" the ambient medium during relaxation, and squeezes it into small-filling-factor filaments, of huge relative mass densities $\rho_j$ :

\hfill{}$\rho_e /\rho_H  =  6 k T / m_H c^2  =  10^{-5.3} T_7$ \hfill{}(14)

\textbf{ }

\noindent (in pressure balance, where X-ray temperatures T have been inserted, via $T_7$). In this process, the pair plasma loses 1/3 of its injected energy, i.e. decelerates significantly, and "entrains" the thermal inclusions at a maximum speed of  $c \rho_e /\rho_H = 10^{-6} c T_7$ , i.e. leaves them practically in their former state of motion. There is no beam beyond the terminating hotspot, yet there is ordered streaming at less than (2/3)c.   

 \textbf{}

\textbf{\hfill{}5.  Realistic Jets\hfill{}}

\textbf{ }

Our above treatment of astrophysical jets -- being scale-invariant, and involving very similar central engines -- is meant to apply to all (hundreds of) known jet sources: from (a) newly forming stars, (b) forming white dwarfs, (c) young neutron stars and BHCs (which latter are held to be neutron stars inside of heavy accretion disks), and from (d) the compact centers of (active) galactic disks; as elaborated in Kundt (1996, 2002, 2004). 

   In all these cases, the central engine is (thought to be) a rotating magnet involving strongly sheared toroidal magnetic fields, whose intermittent reconnections supply abundant relativistic $e^{\pm}$-pair plasma, whose simultaneously emitted LFWs post-accelerate the pairs, and whose equally present thermal radiation cools the pairs, towards an outgoing relativistic Maxwellian and beyond. A deLaval nozzle forms automatically via the huge inertia of the ambient plasma, and the comoving Poynting flux sharpens the particle distribution to a monoenergetic one, and guides the pairs into the channels rammed by earlier generations. Such extremely relativistic beams are stabilized by their comoving Poynting flux, and by the inertia of the ambient plasma. This (generalized) `unified scheme' for all jet sources deviates in part from most other approaches but compares favourably by being more explicit, more uniform, and more stable, and by not invoking (energy-rising) stochastic accelerations. In the sequel, we will comment on a few well-studied sources standing out by their extreme properties.

   The nearest active galactic nucleus is the center of our own galaxy, Sgr A*, the rotation center of the Milky Way disk. Its (unresolved) mass has been determined from the orbits of a few innermost stars as $10^{6.5 \pm 0.1}M_{\odot}$, its spectrum ranges from $10^{8.7}$Hz to beyond 10 TeV, with an integrated luminosity of  $10^{3.7}L_{\odot}$ peaking near GeV energies, and its output shows flares at IR and X-rays with variability timescales of $\la 17$ min (Melia \& Falcke, 2001; Genzel et al, 2003; Mayer-Hasselwander et al, 1998; Aharonian et al, 2004; Kundt, 1990, 2004; Roy \& Rao, 2004). Are we seeing the innermost Galactic disk almost edge-on, whose present-day output reaches us at only $10^{-6.3}$-times its Eddington value? Is its feeding presently throttled by fountain-like evaporation of the innermost disk? In any case, the radio part of the spectrum of Sgr A*, at $\nu \la 10^{13}$Hz, of slope $\alpha := \partial log S_{\nu} / \partial log \nu = 0.3$, signals monoenergetic synchrotron emission at Lorentz factor $\gamma \approx 10^4$, as does the Arc region (Anantharamaiah et al, 1991), whereas its enclosing emitter -- Sgr A East -- radiates a hard power-law spectrum at radio frequencies, as expected for stalled populations.

   An extreme case among extragalactic jet sources is the giant radio galaxy 3C 445, as concerns lossfree transport through large distances. Prieto et al (2002) emphasized the need for in-situ electron acceleration inside the hotspots, based on optical (synchrotron) emission nearly 0.3 Mpc away from the nucleus. From our equ. (11) it is clear that such a problem does not arise for \textbf{E} x \textbf{B}-drifting beams. See also Brunetti (2002), Hardcastle et al (2003), and Stawarz (2004) for similar well-studied sources.

   An even more extreme jet source is the quasar 3C 273, almost unique among  $10^3$ known radio jets by its brightness and one-sidedness (Morrison et al, 1984; Kundt \& Gopal-Krishna, 1986; Jester et al, 2001, 2002). Does its head plough
almost luminally into its CGM ($\beta_h \ga 0.6$), and nearly towards us, with very little resistance in the (cosmic-ray?) halo of its host galaxy? Does its emitted spectrum soften on approach of its tip because of accumulating radiation losses, or because of increased beaming in forward direction (so that we observe a non-representative spectrum)? A rare source may well require a rare explanation.      

   Proceeding to stellar jet sources, there is the unique binary neutron star, or BHC  SS 433, whose interpretation has been controversial ever since its discovery as a jet source, in 1978 (cf. Fender, 2003; Kundt, 1996). Does it emit `bullets' of local-galactic composition, at largely super-Eddington power, or are its beams composed of pair plasma, like in all the other jet sources? Kundt (2004) interprets its moving optical and X-ray emission lines as emitted by the impacted wind matter of its massive companion star, which forms its channel walls, and which is dragged along by the relativistic flow at a fraction of the speed of light. In the latter interpretation, we deal with a $10^4$-year young binary Galactic neutron star inside its SNR W 50, at a distance of 3 Kpc, whose (sub-Eddington) spindown power is still strong enough to prevent accretion from its disk onto its surface, and to blow pairplasma jets whose heads have already crossed the periphery of W 50.

   That jets from Galactic binary neutron stars, or BHCs  may consist of pairplasma has been recently advocated by Kaiser \& Hannikainen (2002), via detections of the redshifted 511 KeV pair-annihilation line in eight binary X-ray sources with jets. When compared with that same emission line from the Crab pulsar (Massaro et al, 1991), its redshift (of some 7\%) is more likely gravitational redshift from a neutron star's surface, where the density of slow pairs should be vastly higher than anywhere downstream along the jet.  Another indication of highly relativistic electrons in compact stellar sources is the superluminal X-ray jet in the microquasar XTE J1550-564, whose observed `deceleration' may have to be understood as a varying phase velocity (Corbel et al, 2002).

   Finally, there are classes (b) (of PNe: Kundt, 1996; Balick \& Frank, 2002), and (a) (of YSOs) which are hard to analyse because enshrouded by dense ionized, atomic, and/or molecular gas and dust. Among the few convincing (nonthermal) YSO candidates are two triple radio sources of (expansion) age $\la 10^3$yr, one of them S 68 in Serpens (Rodr\'iguez et al, 1989), further some 23 often one-sided core radio jets in stellar bipolar flows, including L 1455 (Schwartz et al, 1985), and HH 111 (Reipurth \& Heathcote, 1993;
Rodr\'iguez \& Reipurth, 1994; Reipurth \& Bally, 2001), and the two-sided synchrotron jet from W3(OH) (Wilner et al, 1999). Note that in view of the opacity effects, a reliable distinction between thermal and synchrotron radiation has not always been possible.   

\textbf{ }

\hfill{}\textbf{Acknowledgements\hfill{}}

\textbf{ }

WK owes warm thanks to Christoph Hillemanns, G\"unter Lay, Ole Marggraf, and Christof Wenta for help with the electronic data handling, in connection with this communication, with his homepage, and with saving him from getting drowned by spam mail. Both authors are thankful to Paul Wiita for helpful criticism of an earlier version.

\textbf{}

\textbf{\hfill{}References\hfill{}}

\textbf{ }

Aharonian, F., et al (100 authors) 2004, VHE gamma rays from the direction of Sgr A*, \textit{Astron. \& Astrophys.} \textbf{425}, L13-17.

Anantharamaiah, K.R., Pedlar, A., Ekers, R.D., Goss, W.M.  1991, Radio studies of the Galactic Centre II. The arc, threads and related features at 90 cm (330 MHz),  \textit{MNRAS} \textbf{249}, 262-281.

Balick, B., Frank, A. 2002, Shapes and Shaping of Planetary Nebulae, \textit{Ann. Rev. Astron. \& Astrophys.} \textbf{40}, 439-486.

Baumann, H. 1993, \textit{Astrophysikalische Jet-Phänomene und ihre Deutung als Paarplasma-Strahlen}, Diplomarbeit Bonn.

Begelman, M., Blandford, R.D., Rees, M.J. 1984, Theory of extragalactic radio sources, \textit{Rev. Mod. Phys}. \textbf{56}, 255-351.

Begelman, M., Rees, M.J., Sikora,  1994, Energetic and Radiative Constraints on Highly Relativistic Jets, \textit{Astrophys. J.} \textbf{286}, L57-L60.

Beresnyak, A.R., Istomin, Ya.N., Pariev, V.I. 2003, Relativistic parsec-scale jets: I. Particle acceleration, \textit{Astron. \& Astrophys.} \textbf{403}, 793-804.

Blandford, R.D. 2001, in: \textit{Particles and Fields in Radio Galaxies}, ASP Conference Series \textbf{250}, 487-498.

Blandford, R.D., Rees, M.J. 1974, A `Twin-Exhaust' Model for double Radio Sources, \textit{MNRAS} \textbf{169}, 399-415.

Blome, H.-J., Kundt, W. 1988, Leptonic Jets from Young Stellar Objects?, \textit{Astrophys. \& and Space Sci.} \textbf{148}, 343-361.

Brunetti, G. 2002, Non-thermal Emission from Extragalactic Radio Sources: a High-Resolution - Broad-Band approach, in \textit{The role of VLBI in Astrophysics, Astrometry and Geodesy}, (eds.) F. Mantovani \& A. Kus, Nato Science Ser \textbf{135}, 2004, pp. 29-82. arXiv: astro-ph/0207671.

Corbel, S., Fender, R.P., Tzioumis, A.K., Tomsick, J.A., Orosz, J.A., Miller, J.M., Wijnands, R., Kaaret, P. 2002, Large-Scale, Decelerating, Relativistic X-ray Jets from the Microquasar XTE J1550-564, \textit{Science} \textbf{298}, 196-199.

Eilek, J., Hardee, P., Markovic, T., Ledlow, M., Owen, F. 2002, On dynamical models for radio galaxies, \textit{New Astron. Rev. }\textbf{46}, 327-334.

Fender, R. 2003, Jets from X-ray binaries, \textit{astro-ph}/0303339, 28 August.

Genzel, R., Schödel, R., Ott, T., Eckart, A., Alexander, T., Lacombe, F., Rouan, D., Aschenbach, B. 2003, Near-infrared flares from accreting gas around the supermassive black hole at the Galactic Centre, \textit{Nature} \textbf{425}, 934-937.

Goldreich, P., Julian, W.H. 1969, Pulsar Electrodynamics, \textit{Astrophys. J.} \textbf{157}, 869-880.

Gopal-Krishna, Wiita, P.J. 1988, Hot gaseous coronae of early-type galaxies and their radio luminosity function, \textit{Nature }\textbf{333}, 49-51.

Gopal-Krishna, Wiita, P.J., Hooda, J.S. 1996, Weak headed quasars, \textit{Astron. Astrophys. }\textbf{316}, L13-16.

Hardcastle, M.J., Worrall, D.M., Kraft, R.P., Forman, W.R., Jones, C., Murray, S.S. 2003, Radio and X-ray Observations of the Jet in Centaurus A, \textit{Astrophys. J.}      \textbf{593}, 169-183.

Hones, E.W., Bergeson, J.E. 1965, Electric Field Generated by a Rotating Magnetized Sphere, \textit{J. Geophys. Res.} \textbf{70}, 4951-4958.

Jauncey, D.L., Bignall, H.E., Lovell, J.E.J., Kedziora-Chudczer, L., Tzioumis, A.K., Macquart, J.-P., Rickett, B.J. 2003, Interstellar Scintillation and Radio Intra-Day Variability, in \textit{Radio Astronomy at the Fringe}, eds. J.A. Zensus, M.H. Cohen, \& E. Ros, ASP Conference Series \textbf{300}, 199-210.

Jester, S., Röser, H.-J., Meisenheimer, K., Perley, R., Conway, R. 2001, HST optical spectral index map of the jet of 3C 273, \textit{Astron. \& Astrophys.} \textbf{373}, 447-458.

Jester, S., Röser, H.-J., Meisenheimer, K., Perley, R. 2002, X-rays from the jet of 3C 273: Clues from the radio-optical spectra, \textit{Astron. \& Astrophys.} \textbf{385}, L27-L30.

Kaiser, C.R., Hannikainen, D.C. 2002, Pair Annihilation and Radio Emission from Galactic Jet Sources: the case of Nova Muscae, \textit{MNRAS} \textbf{330}, 225-231.

Katsouleas, T. 2004, Electrons hang ten on laser wake, \textit{Nature} \textbf{431}, 515-516.

Kulsrud, R.M., Ostriker, J.P., Gunn, J.E. 1972, Acceleration of cosmic rays in supernova remnants, \textit{Phys. Rev. Lett.} \textbf{28}, 636-639.

Kundt, W. 1979, A Model for Galactic Centres, \textit{Astrophys. \& Space Sci. }\textbf{62}, 335-345.

Kundt, W. 1986, Particle Acceleration by Pulsars, in: \textit{Cosmic Radiation in Contemporary Astrophysics}, NATO ASI C \textbf{162}, Reidel, 67-78.

Kundt, W. 1990, The Galactic Centre, \textit{Astrophys. and Space Sci.}\textbf{172}, 109-134.

Kundt, W. 1996, in: \textit{Jets from Stars and Galactic Nuclei}, ed. W. Kundt, Lecture Notes in Physics \textbf{471}, 1-18.

Kundt, W. 2002, Radio Galaxies powered by Burning Disks, \textit{New Astronomy Reviews} \textbf{46}, 257-261.

Kundt, W. 2004, \textit{Astrophysics, a new approach}, Springer, Chapter 11.

Kundt, W., Gopal-Krishna 1980, Extremely-relativistic Electron-Positron Twin-jets from Extraga-lactic Radio Sources, \textit{Nature} \textbf{288}, 149-150.

Kundt, W., Gopal-Krishna 1986, The jet of the quasar 3C 273, \textit{J. Astrophys. Astr.} \textbf{7}, 225-236.

Massaro, E. Matt, G., Salvati, M., Costa, E., Mandrou, P., Niel, M., Olive, J.F., Mineo, T., Sacco, B., Scarsi, L., Gerardi, G., Agrinier, B., Barouch, E., Comte, R., Parlier, B., Masnou, J.L. 1991, \textit{Astrophys. J.} \textbf{376}, L11-L14.

Mayer-Hasselwander, H.A., Bertsch, D.L., Dingus, B.L. Eckart, A., Esposito, J.A., Genzel, R., Hartman, R.C., Hunter, S.D., Kanbach, G., Kniffen, D.A., Lin, Y.C., Michelson, P.F., Mücke, A., von Montigny, C., Mukherjee, R., Nolan, P.L., Pohl, M., Reimer, O., Schneid, E.J., Sreekumar, P., Thompson, D.J. 1998, High-Energy gamma-ray emission from the Galactic Center, Astron. \& Astrophys. \textbf{335}, 161-172.

Melia, F., Falcke, H. 2001, The Supermassive Black Hole at the Galactic Center, \textit{Ann. Rev. Astron. \& Astrophys.} \textbf{39}, 403-455.

Michel, F.C. 1974, Rotating Magnetosphere: Acceleration of Plasma from the Surface, \textit{Astroph. J.} \textbf{192}, 713-718.

Morrison, P. 1981, unpublished evening lecture at Socorro, during the IAU \textbf{97} Symposium at Albuquerque on \textit{Extragalactic Radio Sources}.

Morrison, P., Roberts, D., Sadun, A. 1984, Relativistic Jet meets Target: The $\gamma$-rays from 3C 273, \textit{Astroph. J. }\textbf{280}, 483-490.

Ozernoy, L.M., Usov, V.V. 1977, Regular Optical Variability of Quasars and Nuclei of Galaxies as a Clue to the Nature of their Activity, \textit{Astron. Astrophys. }\textbf{56}, 163-172.

Prieto, M.A., Brunetti, G., Mack, K.-H. 2002, Particle Accelerations in the Hot Spots of Radio Galaxy 3C 445, imaged with the VLT, \textit{Science} \textbf{298}, 193-195.

Reipurth, B., Bally, J. 2001, Herbig-Haro Flows: Probes of Early Stellar Evolution, \textit{Ann. Rev. Astron. \& Astrophys.} \textbf{39}, 403-455.

Reipurth, B., Heathcote, S. 1993, Observational Aspects of Herbig-Haro Jets, in: \textit{Astrophysical Jets}, eds. D. Burgarella, M. Livio, \& C.P. O'Dea, Space Tel. Sci. Inst. Symp. \textbf{6}, 35-71.

Rodr\'iguez, L.F., Reipurth, B. 1994, The exciting source of the HH 111 jet complex: VLA detection of a one-sided radio jet, \textit{Astron. \& Astrophys.} \textbf{281}, 882-888.

Rodr\'iguez, L.F., Curiel, S., Moran, J.M., Mirabel, I.F., Roth, M., Garay, G. 1989, Large Proper Motions in the remarkable triple radio source in Serpens, \textit{Astrophys. J.} \textbf{346}, L85-L88.

Roy, S., Rao, A.P. 2004, Sgr A* at low radio frequencies: Giant Metre-wave Radio Telescope Observations, \textit{Mon. Not. R. Astron. Soc.} \textbf{349}, L25-L29.

Scheuer, P.A.G. 1996, Simple Sums on Burning Disks, in: \textit{Jets from Stars and Galactic Nuclei}, Lecture Notes in Physics \textbf{471}, Springer, 35-40.

Schwartz, P.R., Frerking, M.A., Smith, H.A. 1985, Pedestal Features in Dark Clouds: a search for radio emission, \textit{Astrophys. J.} \textbf{295}, 89-93.

Singal, A.K., Gopal-Krishna 1985, Ultra-relativistic bulk motion and radio flux variability of Active Galactic Nuclei, \textit{MNRAS} \textbf{215}, 383-393.

Stawarz, L. 2003, Multifrequency Radiation of Electromagnetic Large-Scale Jets, \textit{Chinese Journal of  A. \& A.}, submitted; astro-ph/0310795 v2.

Wagner, S.J., Witzel, A. 1995, Intraday Variability in Quasars and BL Lac Objects, \textit{Annu. Rev. Astron. Astrophys. }\textbf{33}, 163-197.\textit{ }

Wilner, D.J., Reid, M.J., Menten, K.M. 1999, The synchrotron jet from the $H_2 O$ maser source in W3(OH), \textit{Astrophys. J.} \textbf{513}, 775-779.

\textbf{ }

\end{document}